# Emission line predictions for mock galaxy catalogues: a new differentiable and empirical mapping from DESI

Ashod Khederlarian[1]*, Jeffrey A. Newman,[1] Brett H. Andrews,[1] Biprateep Dey[1], John Moustakas,[2] Andrew Hearin,[3] Stéphanie Juneau,[4] Luca Tortorelli,[5] Daniel Gruen,[5] ChangHoon Hahn,[6] Rebecca E. A. Canning,[7] Jessica Nicole Aguilar,[8] Steven Ahlen,[9] David Brooks,[10] Todd Claybaugh,[8] Axel de la Macorra,[11] Peter Doel,[10] Kevin Fanning,[12,13] Simone Ferraro,[8,14] Jaime Forero-Romero,[15,16] Enrique Gaztañaga,[7,17,18] Satya Gontcho A Gontcho,[8] Robert Kehoe,[19] Theodore Kisner,[8] Anthony Kremin,[8] Andrew Lambert,[8] Martin Landriau,[8] Marc Manera,[20,21] Aaron Meisner,[4] Ramon Miquel,[21,22] Eva-Maria Mueller,[23] Andrea Muñoz-Gutiérrez,[11] Adam Myers,[24] Jundan Nie,[25] Claire Poppett,[8,14,26] Francisco Prada,[27] Mehdi Rezaie,[28] Graziano Rossi,[29] Eusebio Sanchez,[30] Michael Schubnell,[31] Joseph Harry Silber,[8] David Sprayberry,[4] Gregory Tarlé,[31] Benjamin Alan Weaver,[4] Zhimin Zhou[25] and Hu Zou[25]

*Affiliations are listed at the end of the paper*

## ABSTRACT

We present a simple, differentiable method for predicting emission line strengths from rest-frame optical continua using an empirically determined mapping. Extensive work has been done to develop mock galaxy catalogues that include robust predictions for galaxy photometry, but reliably predicting the strengths of emission lines has remained challenging. Our new mapping is a simple neural network implemented using the JAX Python automatic differentiation library. It is trained on Dark Energy Spectroscopic Instrument Early Release data to predict the equivalent widths (EWs) of the eight brightest optical emission lines (including H$\alpha$, H$\beta$, [O II], and [O III]) from a galaxy's rest-frame optical continuum. The predicted EW distributions are consistent with the observed ones when noise is accounted for, and we find Spearman's rank correlation coefficient $\rho_s >$ 0.87 between predictions and observations for most lines. Using a non-linear dimensionality reduction technique, we show that this is true for galaxies across the full range of observed spectral energy distributions. In addition, we find that adding measurement uncertainties to the predicted line strengths is essential for reproducing the distribution of observed line-ratios in the BPT diagram. Our trained network can easily be incorporated into a differentiable stellar population synthesis pipeline without hindering differentiability or scalability with GPUs. A synthetic catalogue generated with such a pipeline can be used to characterize and account for biases in the spectroscopic training sets used for training and calibration of photo-$z$'s, improving the modelling of systematic incompleteness for the Rubin Observatory LSST and other surveys.

**Key words:** methods: data analysis – methods: numerical – galaxies: ISM – galaxies: stellar contents.

## 1 INTRODUCTION

The next generation of deep, wide-field photometric surveys (Stage IV, see Albrecht et al. 2006) coming online this decade (including LSST, Ivezić et al. 2019; Roman, Spergel et al. 2015, and Euclid, Laureijs et al. 2011) will strongly constrain cosmology by probing dark matter and dark energy using a variety of methods. To take full advantage of these new data sets, stringent requirements have been set on both the performance of photometric redshifts (photo-$z$'s) for individual objects and the characterization of photo-$z$ distributions for groups of objects (Mandelbaum et al. 2018). State-of-the-art photo-$z$ algorithms do not meet these requirements, partly due to our incomplete knowledge of the galaxy population (see Newman & Gruen 2022 for a review).

To minimize the impact of incorrect redshift measurements on cosmology studies, spectroscopic data sets used for calibrating photo-$z$ algorithms will need to be restricted to only galaxies with highly-confident redshift measurements (Newman & Gruen 2022). However, such samples are biased towards galaxies with strong spectral features such as emission lines. As a result, regions in colour–magnitude space that lack such features will be systematically underrepresented in spectroscopic samples, and photo-$z$ algorithms will have to extrapolate over them. This is especially a problem at greater depths, where the fraction of highly secure redshifts in spectroscopic surveys such as DEEP2 (Newman et al. 2013) and zCOSMOS (Lilly et al. 2007) becomes small (Newman et al. 2015).

* E-mail: ask126@pitt.edu

Using a simple mock catalogue, Hartley et al. (2020) showed that limiting the spectroscopic training set to objects with confident redshifts can significantly bias the predicted redshift distribution for a group of galaxies even in current surveys, resulting in systematic errors that are significantly greater than the upper limits set for LSST weak lensing tomographic bins (Mandelbaum et al. 2018). Characterising and accounting for this systematic will be necessary to achieve the ambitious science goals set for Stage IV surveys. A mock galaxy catalogue that realistically models galaxy spectral energy distributions (SEDs) and the incompleteness in spectroscopic data sets could enable this. However, this will require accurate modelling of the features used to measure redshifts, particularly the strongest emission lines.

Such catalogues can be generated starting from darkmatter-only simulations (Villaescusa-Navarro et al. 2020) and populating haloes at each epoch with galaxies using the galaxy–halo connection, which is constrained by both galaxy clustering and weak gravitational lensing measurements (Wechsler & Tinker 2018). In addition to physical properties at a certain epoch, galaxy accretion histories and star formation rates (SFRs) across cosmic time can also be modelled from the accretion histories of their host haloes (Behroozi et al. 2019; Alarcon et al. 2023). The correlation between galaxy assembly and halo assembly is constrained by measurements such as specific and cosmic SFRs (Popesso et al. 2023) and stellar mass functions (Moustakas et al. 2013).

The history of a galaxy's star formation can then be used to predict its spectrum, which will be an amalgamation of the stellar continuum generated by its stars (or an active galactic nucleus) and the interaction of this continuum with the surrounding gas and dust. Synthetic stellar continua are typically generated using stellar population synthesis (SPS) models (Bruzual & Charlot 2003; Conroy & Gunn 2010; Conroy 2013) or by combining empirical templates (Connolly et al. 1994; Kinney et al. 1996; Brown et al. 2014). Both approaches have been used to forward-model realistic galaxy populations with precise redshift distributions (Tortorelli et al. 2021; Alsing et al. 2023; Moser et al. 2024). In general, these predictions also depend on dust attenuation laws (e.g. Salim, Boquien & Lee 2018).

Emission from gas includes both nebular continuum emission and nebular emission lines. The former corresponds to continuous emission from free–free, free–bound, and two-photon emission, while the latter is emission at specific wavelengths generated by recombination processes and line transitions. Emission in H II regions can be predicted using photoionization codes (Ferland et al. 2017; Jin, Kewley & Sutherland 2022); however, this involves making simplifying assumptions about the structure and composition of the gas to reduce the number of free parameters and make the problem tractable (Byler et al. 2017). If our purpose is only to predict realistic emission line strengths without consideration of the physical properties driving those fluxes, making such assumptions can be avoided by empirically mapping from continua to emission lines. In addition, such a mapping can account for contributions from active galactic nuclei (AGNs) which impact the continuum and the strengths of emission lines in a correlated way. The focus of this paper is to learn said mapping from observed spectra in a way that facilitates applications on synthetic data.

Understanding the correlation between emission lines and continua is also important for interpreting observations. In young simple stellar populations (SSPs) with sub-solar metallicity, nebular emission can contribute significantly to broad band fluxes, even up to 60 per cent in extreme cases (Anders 2003). The emission lines have the greatest impact on optical fluxes, while nebular continuum emission becomes more significant in the near infrared (Byler et al. 2017). This is most apparent in young massive star clusters, where some broad-band colours and magnitudes cannot be interpreted without accounting for nebular emission, and ignoring it in modelling can also significantly affect inferred properties such as age and mass (Anders 2003; Reines et al. 2009).

Galaxies are composite stellar populations, so those that have a substantial population of young stars will exhibit strong contributions from nebular emissions to broad-band fluxes. This can lead to contamination in colour-selected galaxies (Schaerer & de Barros 2009; Atek et al. 2011), and it has also been shown that incorporating the impact of emission lines on observed colours can improve template-based photo-$z$ estimates (Győry et al. 2011). More recent observations with *JWST* (Gardner et al. 2006) also highlight the importance of emission line modelling in interpreting galaxy SEDs (Naidu et al. 2022). For inferring the properties of a population of galaxies, Alsing et al. 2024 (see also Leistedt et al. 2023) have used an empirical emission line re-calibration scheme to properly account for contributions to photometry.

The existence of a mapping from continua to emission lines is supported by the observed correlations between the physical parameters that drive both stellar and nebular emission, such as the mass–metallicity (Tremonti et al. 2004; Andrews & Martini 2013) and mass–SFR (Popesso et al. 2023) relations. Such relations indicate that the stellar continuum should also contain information about the same physical properties that affect emission line fluxes, such as the history of a galaxy's formation of stars and metals. However, it is important to note that, given the observed scatter in these relations, we should also expect some intrinsic scatter in any mapping between continuum and emission line properties. This can be caused by observational uncertainties, non-homogeneous dust attenuation, and bursts of star formation.

Currently, the aforementioned mapping can only be inferred from low-redshift galaxies, since high signal-to-noise and high-resolution rest-frame optical spectra are required. This then raises the question of applying it to add emission lines on synthetic continua of higher-redshift galaxies; for characterizing spectroscopic incompleteness in training sets used for e.g. lensing analyses, realistic emission lines are required for redshifts up to $z \sim 1.2$ (Mandelbaum et al. 2018). Juneau et al. (2014) suggest that some of the observed evolution in emission line ratios can be explained by selection effects. Other observations and simulations suggest that redshift evolution of optical emission line strengths (and strengths of ratios) is correlated, and possibly caused by, higher ionization parameters (due to higher SFRs) and lower-metallicities, both of which are also reflected in continuum shapes (Kewley et al. 2015; Hirschmann et al. 2023; Backhaus et al. 2024). Therefore, to the extent that high-redshift galaxies are well-represented (via low-redshift analogs) in the training set used to learn the mapping, extrapolating to higher redshifts should be possible. In Section 5.2 we discuss one possible way of testing this hypothesis, which involves using spectra of low-redshift extremely metal-poor dwarf galaxies (Zou et al. 2024).

Recent efforts to characterize the relationship between continua and emission lines have focused on spectra from the Sloan Digital Sky Survey (SDSS; York et al. 2000). Using principal component analysis (PCA), Győry et al. (2011) showed that strong correlations exist between stellar continua and emission lines. Beck et al. (2016) took this a step further by using local weighted linear regression (LWLR) to predict emission line equivalent widths (EWs) from the continuum PCA coefficients. However, in that study the training set was limited to high signal-to-noise spectra for which all emission lines of interest were required to be non-zero, limiting this effort to strongly star-forming galaxies or AGN. Also, because LWLR relies on finding nearest neighbours in the space of input variables ('features', in machine learning parlance), it is neither differentiable nor scalable; the lack of these advantages impedes the integration of such methods with state-of-the-art SPS codes such as differentiable stellar population synthesis (DSPS, Hearin et al. 2023).

In this paper, we focus on developing a new approach to this problem that can be easily integrated with DSPS. Since SPS models are the main bottleneck for both forward-modelling galaxy populations (Alsing et al. 2020, 2023) and for inferring galaxy properties from observations (Johnson et al. 2021), DSPS offers a fast and scalable alternative. It is implemented in a software library that supports automatic differentiation (JAX; Bradbury et al. 2018), making it differentiable with respect to its input parameters. This allows the use of gradient-based inference methods such as Adam optimization (Kingma & Ba 2014) and Hamiltonian Monte Carlo (Duane et al. 1987). It is also scalable because JAX functions can easily be ported on to GPUs. Compared to standard SPS codes, DSPS provides speedups of a factor of $\sim 5$ on a CPU and a factor of 300–400 on a modern GPU.

Currently, DSPS can incorporate photoionization-based emission lines by employing SSP templates from Byler et al. (2017) that include nebular emission. Adding alternative empirically predicted emission lines to DSPS synthetic stellar continua would be best done via an approach that maintains the scalability and differentiability of DSPS.

Having these goals in mind, we present a simple neural network implemented in JAX which predicts the EWs of eight strong optical emission lines ([O II] doublet, H$\gamma$, H$\beta$, [O III]$\lambda$5007, H$\alpha$, [N II]$\lambda$6584, [S II]$\lambda$6716, and [S II]$\lambda$6731) given a continuum. To train the network, we use Early Release data from the Dark Energy Spectroscopic Instrument (DESI, Collaboration 2023), including objects with arbitrarily small emission-line strengths to limit bias. We directly compare the effectiveness of our techniques to the PCA/LWLR method employed by Beck et al. (2016) by applying both to the same data set, and we also explore predicting EWs from galaxy parameters inferred from the continua.

$$\lambda \qquad (1)$$

In general, for a given continuum, one would expect a probability distribution of EW values $p(EW|continuum)$ that would capture observational uncertainties, intrinsic scatter in the continuum–emission line relation, and even covariances between different line strengths. However, given our principal motivation of modelling spectroscopic incompleteness, we are primarily concerned with estimating whether an emission line is strong enough to yield a successful redshift measurement. Therefore, a simple deterministic neural network that predicts a point estimate of $p(EW|continuum)$ is sufficient for our purposes. Several approaches can potentially be used to obtain EW probability distributions (e.g. Bayesian neural networks Goan & Fookes 2020 and normalizing flows Kobyzev, Prince & Brubaker 2020) and to ensure that they are conditionally calibrated (Dey et al. 2022). We leave this for future work as it is beyond the scope of this paper.

The structure of the paper is as follows. In Section 2, we describe the DESI dataset; Sections 2.2 and 2.4 elaborate on how the continuum and emission line EWs are obtained from the observed spectra. Section 3 describes our fiducial JAX-neural-network method and the two methods it is compared to. In Section 4, we assess performance by comparing predicted and observed EW distributions (Section 4.1), reproducing line-ratio diagnostic diagrams (Section 4.2), and using Uniform Manifold Approximation and Projection (UMAP; McInnes et al. 2018b) embeddings to analyse across different galaxy SEDs (Section 4.3). Finally, we conclude in Section 5 with a summary and discussion of future work. Throughout this paper we use standard flat $\Lambda$CDM cosmology with present-day matter density parameter $\Omega_m = 0.3$ and Hubble constant $H_0 = 70 \text{kms}^{-1}\text{Mpc}^{-1}$.

## 2 DATA

In this section, we elaborate on how DESI data sets, in particular from the Bright Galaxy Survey (BGS; Hahn et al. 2023b, Juneau et al. 2024), were employed in our analyses and how they were prepared for use in machine learning algorithms. To limit biases, we made minimal quality cuts to remove objects that are not classified as galaxies and to retain only spectra that have confident redshift measurements. With the goal of using our method to add emission lines on synthetic continua, we emphasized obtaining representations of the observed DESI spectra that are not sensitive to features which might not be present in the mock data set.

### 2.1 DESI and the DESI Bright Galaxy Survey

DESI is a 5000-fiber spectrograph installed on the Mayall 4m telescope at Kitt Peak National Observatory (DESI Collaboration 2022). Over the course of its operations, it will obtain spectra in the wavelength range 3600–9800 Å (with spectral resolution between 2000 and 5500) for over 40 million galaxies and quasars over an area of at least 14 000 deg$^2$ (DESI Collaboration 2016b; Miller et al. 2023; Silber et al. 2023). It is the first Stage IV experiment for probing the nature of dark energy to begin operations (Levi et al. 2013; DESI Collaboration 2016a).

All DESI targets are selected using photometric catalogues derived from the DESI Legacy Imaging Surveys (Dey et al. 2019), which covered more than 14 000 deg$^2$ of sky in the $g$ (4700 Å), $r$ (6230 Å), and $z$ (9130 Å) optical bands. The catalogue also contains model-matched near- and mid-infrared photometry from the WISE satellite and NEOWISE-Reactivation (Wright et al. 2010; Mainzer et al. 2014). Photometry for large galaxies in the local universe was handled differently in the Siena Galaxy Atlas (Moustakas et al. 2023b).

In order to train and test the machine learning algorithms employed in this work, we require a broad sample of bright galaxies at low redshift whose spectra overlap in the rest-frame optical and that span a wide range of galaxy properties. The DESI BGS Bright target class (Hahn et al. 2023b) fits our needs; it is selected using only an $r < 19.5$ magnitude limit, resulting in an expected target density > 800 targets/deg$^2$, and spanning the redshift range of $0 < z < 0.6$. DESI Survey Validation observations (DESI Collaboration 2023a) showed that BGS Bright targets meet their science requirements, with a target density of $\sim 860$ targets deg$^{-2}$, > 80 per cent fiber assignment rate (meaning spectra are obtained for > 80 per cent of potential targets), > 95 per cent redshift success rate for assigned fibers, and < 1 per cent stellar contamination. The redshift success rate does vary in colour–magnitude space, but it is > 95 per cent within most regions (stays > 90 per cent throughout), meaning successful redshifts are obtained for a broad range of galaxies. This makes the BGS Bright sample ideal for our purposes, as opposed to other data sets that have been used in previous work; in particular, Beck et al. (2016) used the main Bright sample from SDSS Data Release Seven (Abazajian et al. 2009), which is selected with Petrosian $r$-band magnitude <17.77, limiting the diversity of low-mass, faint galaxies. In addition, DESI boasts a significantly higher instrument throughput, wider wavelength coverage, higher spectral resolution, larger mirror, and better flux calibration. These advantages allow DESI to achieve a comparable signal-to-noise ratio (S/N) with significantly shorter exposure times, which is one of the reasons why it will obtain at least an order of magnitude more spectra by the end of operations.

### 2.1.1 Data processing pipelines

The DESI spectrographs incorporate three cameras (blue, red, and NIR), resulting in three spectra for each target. In cases where targets were observed multiple times, the spectra for each object were coadded to provide a single spectrum per camera. As a final processing step we coadd the blue, red, and NIR spectra for each object (which overlap in wavelength coverage), resulting in a single spectrum covering the full wavelength range [3600–9800]Å. Flux calibration is performed by fitting the spectra of standard stars with model templates; based on tests with white dwarf spectra, calibration is typically good to ±2 per cent residuals, with larger deviations (∼ ±6 per cent) at the bluest end. (Guy et al. 2023) provides a detailed overview of the DESI spectroscopic data-processing pipelines.

Redshifts are measured from spectra using `Redrock`, which selects the best classification (galaxy, quasar, or star) and redshift for an object based upon minimizing the $\chi^2$ difference between spectral templates and the observed spectrum. The spectroscopic data processing pipelines were validated by visual inspections of a subset of objects, as described in (Alexander et al. 2023; Lan et al. 2023). We also make use of two value-added catalogues: `FastSpecFit` (Section 2.2), and `PROVABGS` (Section 3.2.2).

### 2.1.2 The DESI Early Data release

The data obtained for DESI Survey Validation (SV) were recently publicly released (DESI Collaboration 2023). The resulting sample includes ∼285 000 spectra obtained to test the BGS Bright survey design. SV was split into two main phases: Target Selection Validation (SV1) and the One-Percent Survey (SV3) (DESI Collaboration 2023a). The SV1 targets extended beyond the main survey selection boundaries and were observed for longer than the nominal exposure time of 180s; the resulting sample was used to finalize the DESI operations program as well as target selection. SV3 targets were selected to include DESI-like samples for all target classes over one per cent (140 deg$^2$) of the final survey's footprint. The SV3 spectra have been used to test science analysis pipelines and the efficiency of automated routines.

In this paper, we have used SV1 data for both training and validation because the spectra have high *S/N* (they were observed for four times longer than the nominal effective exposure time of the actual survey). The SV3 sample was then used as our test set. The SV1 and SV3 BGS samples have similar distributions in colour, magnitude, and redshift, but SV3 was observed for only the nominal DESI exposure time, resulting in a very different noise distribution than SV1. By training on SV1 and testing on SV3, we can coarsely test whether the noise distribution of a data set (which would be very different in simulated data) has a significant impact on the performance of our methods. We have also tested our methods using equivalent data sets for training and testing (e.g. the combined SV1 + SV3 dataset for both) and obtain results that are as good or better than those reported here in that case.

## 2.2 Emission lines from `FastSpecFit`

Throughout this paper, we use a value-added catalogue produced via the `FastSpecFit` code (Moustakas et al. 2023a) to determine the equivalent widths of emission lines (in particular, we use data release v3.1 that was obtained by running the `FastSpecFit` code v2.4.3). `FastSpecFit` models both stellar continuum and emission lines using methods optimized for both speed and simplicity.[1]

To minimize the impact of reddening, we focus on predictions of emission line EWs, which measure the strengths of the emission lines relative to the continuum at similar wavelength, rather than fluxes. The fluxes for each emission line are first determined by integrating a Gaussian profile fitted to a continuum-subtracted spectrum. From this, EWs are then calculated by dividing the fluxes by the continuum at the line's central wavelength (defined as the median continuum flux within three line widths).

The `FastSpecFit` catalogue also provides estimates of the uncertainties in each EW measurement. In order to speed computation the catalogue uncertainties are obtained by propagating flux errors in the original spectrum, using the best-fittng Gaussian profile to weight them. As a result they do not include contributions from uncertainties in the line profile width, which can be significant at low EWs. For this reason, we use a more conservative estimate of the line flux uncertainties, obtained by summing in quadrature the estimated errors for each pixel that contributes to the emission line (corresponding to the error in the total line flux without weighting pixels using a line profile).

## 2.3 Sample selection

To ensure the inclusion of as broad as set of galaxies as possible, we applied only a minimal set of data quality cuts to the BGS Bright samples, restricting to objects which:

(i) are classified as 'GALAXY' type by `Redrock` (as opposed to 'QSO' or 'STAR');

(ii) have $\Delta\chi^2 \geq 25$ between the best and next-best redshift solution, in order to ensure a secure redshift measurement;

(iii) are in the redshift range $0.05 < z < 0.3$, so that their spectra overlap over the rest-frame wavelength range of [3400-7000]Å;

(iv) and have `Fastspecfit` EW>0 and non-zero EW inverse variance for whichever line is being predicted at a given time.

The first selection cut removes objects classified as quasars or stars; however a small number of AGN remain after the cut (see Section 4.3), in addition to a notable population of galaxies with low-ionization nuclear emission regions (LINERs). See Juneau et al. 2024 for a detailed study of the AGN sample in BGS. A significant portion of emission from LINERs can be attributed to diffuse stellar sources, most likely post-asymptotic giant branch (post-AGB) stars (Singh et al. 2013; Belfiore et al. 2016; Byler et al. 2019).

To minimize any biases towards the bluest, highest star formation rate galaxies, the last selection is applied only on the line that is being predicted; for example, when predicting the EW of H$\alpha$, the training and test sets require EW(H$\alpha$) > 0 and finite EW(H$\alpha$) variance, with no additional constraints on the EWs of other lines. This choice results in eight training and eight test sets, one pair for each line. The redshift, $g - r$ apparent colour, and $r$ apparent magnitude distributions of the SV1 training set for H$\alpha$ is shown in Fig. 1. As mentioned in Section 2.1.2, the SV3 test set has similar distributions. The figure also shows that selecting only galaxies for which all lines are detected would have strong effects on the selection of the sample, mainly biasing it towards bluer galaxies.

---

[1] https://fastspecfit.readthedocs.io/en/latest/#

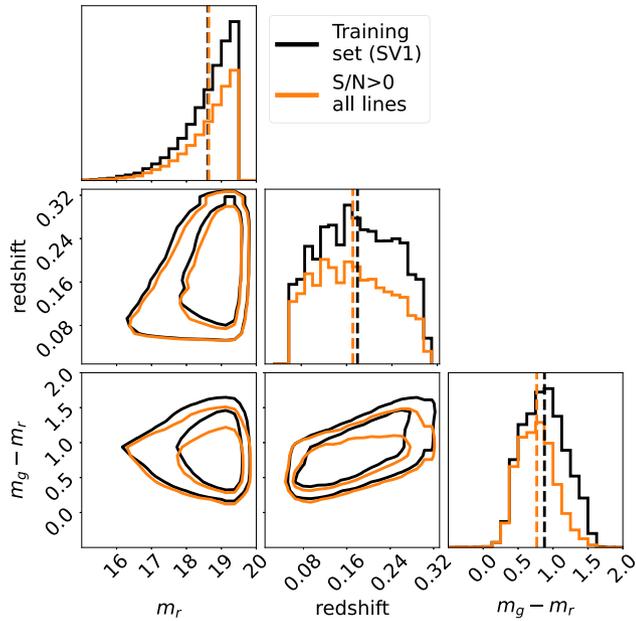

**Figure 1.** Distributions of the 2D projections of apparent $g - r$ colour, apparent $r$ magnitude, and redshift for those objects in the training subset of the DESI SV1 sample that have detections of the H$\alpha$ emission line, with S/N(H$\alpha$) > 0 (black), as well as the corresponding distributions for objects where all eight optical emission lines considered in this paper (Table 1) are also required to have $S/N > 0$ (orange). The magnitudes plotted were obtained from the DESI Legacy Imaging Survey. The orange set includes significantly fewer red galaxies; even more severe restrictive cuts that have been used in the literature would result in a stronger bias.

After the first three cuts (on template type, $\Delta\chi^2$, and redshift) we are left with 36 219 objects from SV1 and 108 709 objects from SV3. We then applied the last cut on each of these parent samples, resulting in ≈27 000–33 000 objects used for training from SV1 and ≈70 000–100 000 objects used for testing from SV3 (with exact numbers differing for each line). Many objects overlap between the training and test sets for different lines (e.g. an object that satisfies the line detection criterion for all lines of interest will of course appear in all the data sets).

After the emission line cuts (EW>0 and finite EW variance), we split the SV1 samples into a training set consisting of 75 per cent of the objects and a validation set consisting of 25 per cent. Similarly, we split the SV3 sample into 50 per cent testing and 50 per cent blind-testing subsets. The validation set was used to keep track of the validation loss during training, and half of SV3 was used to optimise hyperparameters, with the remaining half (the blind test set) being used only for the results shown in this paper. We chose this approach because, as explained in Section 2.1.2, the SV1 and SV3 sets have different noise distributions. For plots of line-ratio diagnostic diagrams, we use the 34 102 objects in common between the blind test sets for H$\beta$, [O III]$\lambda$5007, H$\alpha$, and [N II]$\lambda$6584, and 31 248 objects common between H$\beta$, [O III]$\lambda$5007, H$\alpha$, and [S II]$\lambda$6584.

### 2.4 Continuum measurements

We wish to predict emission line EWs based upon the shape of a galaxy's continuum. To estimate these quantities we first shift each spectrum to the rest-frame and then mask the emission lines of concern over the wavelength windows that are listed in Table 1. We replace the observed flux in the masked regions by a linear interpolation between the smoothed continuum level on either side; this smoothing is performed using an 11-pixel-wide median filter. In addition to the eight emission lines mentioned in the table, the [N II] line at 6548Å is masked with the H$\alpha$ window, and we also mask the [O III] line at 4959 Å using a masking window of size 14 Å.

It is possible to predict the strengths of emission lines directly from the detailed rest-frame continuum of an object by using a convolutional neural network with an attention mechanism, which has been shown by Melchior et al. (2022) to be a good architecture for working with spectra. However, since our ultimate goal is adding emission lines to synthetic stellar continua, we have opted instead to work with a method that is independent of spectral resolution and not sensitive to features in the observed spectra that might not be present in synthetic ones.

Therefore, we instead estimate the average continuum flux within each of $N$ synthetic medium bands evenly spaced in wavelength (linear spacing), each represented by a top-hat filter (in combination covering the full rest-frame wavelength range where our spectra overlap, [3400–7000]Å). We have tested our methods using a variety of values for $N$ (6, 12, 15, 20, 30, 40, and 50). When fewer bins are used we obtain higher-S/N measurements but have a lower effective spectral resolution. The summary statistics we have used to assess performance (described in Section 4) all improve significantly from 6 to 12 bins but plateau at larger $N$. Furthermore, line-ratio diagnostic diagram reconstructions (also explained in Section 4) worsened with increasing $N$. Presumably, the gain in information from using more bins is offset by the decrease in S/N and by the decrease of density in feature space (due to the curse of dimensionality; Ivezić et al. 2020). For these reasons, we adopt $N = 12$ for all results shown in this paper.

More concretely, we calculated the mean flux density in wavelength units over a set of wavelength windows:

$$\langle f_\lambda \rangle_i = \frac{\int_{\lambda_i}^{\lambda_{i+1}} f_\lambda(\lambda) d\lambda}{\int_{\lambda_i}^{\lambda_{i+1}} d\lambda} \quad \text{with} \quad i = 1, \ldots, 12. \quad (2)$$

The wavelength windows correspond to the $N = 12$ synthetic bands defined by the edges $\lambda_i$, with $\lambda_1 = 3400$ Å and $\lambda_{13} = 7000$ Å. The flux density in wavelength units is $f_\lambda(\lambda)$, and its average value in band $i$ is given by $\langle f_\lambda \rangle_i$. We then calculate the 11 log flux ratios between successive bins ($c_i$):

$$c_i = \log\left(\frac{\langle f_\lambda \rangle_{i+1}}{\langle f_\lambda \rangle_i}\right) \quad \text{with} \quad i = 1, \ldots, 11. \quad (3)$$

We used these ratios as inputs (features) for predicting emission line EWs, in combination with a measure of luminosity calculated from the average flux in the bin nearest to 6250Å:

$$L = \log\left[\langle f_\lambda \rangle_{10} \times D_L(z)^2\right], \quad (4)$$

where $D_L(z)$ is the luminosity distance of an object at redshift $z$. The flux ratios provide information on the (normalized) star formation history of a galaxy, which is to first order what determines the EWs. We have found that also incorporating a measure of luminosity (as a proxy for mass at fixed continuum shape) significantly improved predictions for the equivalent width of the [N II]$\lambda$6584 line, presumably because of the observed correlation of nitrogen abundance with stellar mass (Andrews & Martini 2013). A sample spectrum from the training set is shown in Fig. 2, along with the estimates of its binned continuum.

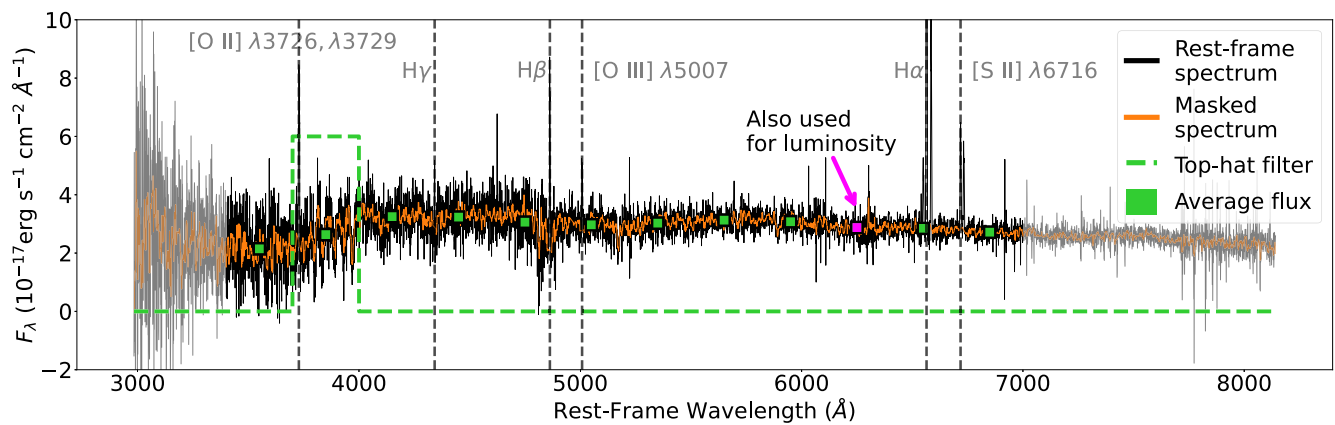

**Figure 2.** An illustration of how the continuum shape is obtained from the observed spectrum. After correcting for Milky Way extinction, the emission lines in the observed spectrum (black) are masked and replaced by a linear interpolation of the median-smoothed continuum (orange). Then, we average the flux in 12 evenly spaced medium bands each described by a top-hat filter; one example filter is shown by the green dashed lines. The resulting continuum measurements (indicated by the green and purple squares) cover the wavelength range [3400–7000]Å (where spectra overlap across our full $0.05 < z < 0.3$ redshift range). We use the set of continuum flux ratios in successive bands, in combination with the luminosity inferred from the filter closest to 6250 Å (purple square), as features for predicting emission line EWs.

## 3 METHODS

Having obtained estimates of both the continuum fluxes and emission line EWs, we now describe our fiducial method for mapping between the two. We have also explored alternative approaches for doing so, even using different representation of the continuum. Given an efficient representation of the latter and a sufficiently complex function to map to EWs, any method can yield satisfying results irrespective of the details; however, these details matter when considering computational efficiency and scalability, which is why we chose a JAX-implemented simple neural network as our fiducial method.

### 3.1 JAX neural network (JAX-NN)

Our neural network is completely implemented in a Python automatic differentiation library called JAX (Bradbury et al. 2018). With this approach, our method is differentiable and GPU-scalable, making it straightforward to integrate with DSPS (which is also implemented in JAX; Hearin et al. 2023).

The network is composed of three hidden layers with 64, 128, and 64 neurons. As described in Section 2.4, the inputs to the network (representing stellar continua) are 11 flux ratios and a luminosity around 6250Å, and the outputs are emission line EWs. Since the EWs were allowed to be arbitrarily small, we found that predicting arcsinh(EW), which behaves logarithmically at large EW values but is finite and linear at the origin, gave the best results. The network was trained to minimize the mean-square-error loss function:

$$L = \frac{1}{N} \sum_{i=1}^{N} \left[ \text{arcsinh}(\text{EW})_i^{\text{pred}} - \text{arcsinh}(\text{EW})_i^{\text{true}} \right]^2. \quad (5)$$

Applying this same loss function but using linear EW values rather than arcsinh(EW) values would give significantly more weight to objects with larger EWs, since the same fractional difference between predicted and observed values will result in a much larger loss when EW is large. If, instead, ln (EW) was used to calculate losses, then the same fractional difference between predicted and observed values when EW $\approx$ 0.1 and EW $\approx$ 100 would have equal weight. However, for our purposes small errors in EW when EW itself is very small do not matter, as it would not affect whether a given emission line would be detected or not in observations when the EW value is negligible in either case. Using an inverse hyperbolic sine function to calculate losses mitigates both of these undesirable behaviours.

For training, we used the Adam optimizer (Kingma & Ba 2014) with parameters $\beta_1 = 0.9$, $\beta_2 = 0.99$, and learning rate $10^{-4}$. The input features ($c_i$ and $L$) were standardized to have a mean of 0 and a standard deviation of 1. The training samples of $\approx 27\,000 - 33\,000$ objects (with the exact number depending upon the line being predicted) were split into 75–25 per cent training-validation samples, as described in Section 2.3. The optimizer was run with a batch size of 2048 and training was stopped when the difference between the validation loss at the current epoch and the average of the losses of the previous 20 epochs was less than $10^{-3}$. Hyperparameter tuning was done using 50 per cent of the SV3 test sets, leaving the remaining 50 per cent for a blind test set used to produce all results in this paper.

### 3.2 Alternative approaches

We have also explored several alternative approaches of mapping stellar continua to emission lines. These alternatives do not have the same advantages of scalability and differentiability that our primary algorithm does; however, they do aid in assessing whether better performance (i.e. better agreement with the test data) could be obtained by following a very different approach.

Among the tested methods, we will only elaborate on the PCA approach of Beck et al. (2016) for the sake of comparison, and on our attempt to predict EWs directly from a set of estimated physical parameters for each object (e.g. stellar mass and parameters that describe SFH for each galaxy), which is a way of estimating emission lines without first synthesizing continua. Other algorithms that we tested but will not describe include highly effective decision tree algorithms, such as XGBoost (Chen & Guestrin 2016), and convolutional neural networks with an attention mechanism. None of these alternatives (including PCA) performed better than our fiducial JAX-NN method.

*3.2.1 Comparing with previous work (PCA-LWLR)*

We applied the method of mapping stellar continua to emission lines in Beck et al. (2016) to provide a literature-based baseline for comparison (despite this method not being differentiable or scalable). This involved first putting the full-resolution stellar continua (orange in Fig. 2), which were extracted from DESI observations, on a common wavelength grid ([3400–7000]Å with 0.8Å grid-spacing) using nearest neighbour interpolation. Then, they were normalized to have the same flux in the rest-frame DECam *g* band (Flaugher et al. 2015) and the average continuum was subtracted from each spectrum. PCA (from Scikit-learn, Pedregosa et al. 2011) was used on the resulting continua and the first five PCA coefficients were kept as a low-dimensional representation (5D-PCA) of each galaxy's continuum.

Emission line EWs were predicted from the 5D-PCA space using LWLR and a training set with known EWs. Given the five PCA coefficients of a galaxy from the test set, the method works by using some distance metric to find *k*-nearest neighbours in this space that belong to the training set. These neighbours are then given certain weights and used to perform weighted linear regression to predict EWs for the test-set galaxy. For another galaxy with unknown EWs, a new set of *k*-nearest neighbours are found and the process is repeated; this series of local weighted linear regressions results in a globally non-linear fit. With our data set, we used the Euclidean distance metric, with $k = 800$, and weights corresponding to inverse distance.

We emphasize that this method can not be implemented in JAX and is therefore neither differentiable nor scalable with GPUs, mainly due to the neighbourhood-search algorithm that is employed.

*3.2.2 EWs from physical parameters (PROVABGS)*

As a third approach, we also attempted to predict EWs directly from a set of estimated physical parameters for each galaxy. Posterior distributions for a variety of galaxy properties are available for BGS SV3 targets in the PROVABGS catalogue (Hahn et al. 2023a). These distributions were determined by jointly modelling the spectroscopy and photometry for each object in a Bayesian framework, using an SPS model that is generated from SSPs calculated with non-parametric star formation histories (SFHs) combined with a short-duration starburst component, non-parametric chemical enrichment histories (ZH), and a two-component dust attenuation model with birth cloud and diffuse-dust components. This results in a total of 13 parameters: stellar mass ($\log M_*$); four coefficients ($\beta_i$ $i = 1, 4$) to express SFH as a linear combination of four basis functions; fraction of total stellar mass formed during the starburst ($f_{\text{burst}}$); time at which the starburst occurs ($t_{\text{burst}}$); two coefficients ($\gamma_1^{ZH}, \gamma_2^{ZH}$) to express ZH in terms of basis functions; birth cloud optical depth ($\tau_{\text{BC}}$); diffuse dust optical depth ($\tau_{\text{ISM}}$); dust index ($n_{\text{dust}}$); and a normalization factor to account for fiber aperture effects ($f_{\text{fiber}}$). Although we do not use the model templates explicitly, we attempt to predict line EWs from 12 of these parameters, excluding $f_{\text{fiber}}$ as it contains information on redshift.

Since PROVABGS values were only available for the SV3 sample, we split our test set of ≈70 000–100 000 objects into a 52.5–17.5 per cent-30 per cent train-validation-test split only for this method. We found that using a tree-based algorithm called XGBoost (Chen & Guestrin 2016) performed better than a JAX-based neural network when predicting EWs from these physical parameters, so we present XGBoost-based results for PROVABGS. We implemented it using the XGBoost Python package[2] with early stopping rounds parameter set to five, learning rate set to 0.05, and maximum tree depth set to nine.

Again, we emphasize that it is not straightforward to implement XGBoost in a differentiable fashion. In addition, a mapping trained from physical parameters is not likely to be broadly applicable to all simulated data sets, as the values for these parameters are inferred from the observed continuum assuming some SPS model that may not be consistent with the SPS used for a set of simulated spectra. PROVABGS also does not include AGN templates, so it can not be used when there are contributions from AGN.

## 4 RESULTS AND ANALYSIS

As described above, we generally have used a common subset of the DESI SV3 BGS sample as a blind test set to assess the performance of each algorithm (with the exception of the PROVABGS-based predictions; cf. Section 3.2.2). In this section, we present the results of these tests. First, we evaluate the overall accuracy of predictions in a more traditional sense by using a set of summary statistics, scatter plots of predicted versus observed EWs, and comparisons of the cumulative EW distributions for each line. Next, to examine how well line ratios are preserved, we test the reproduction of line-ratio diagnostic diagrams. Finally, we use a non-linear dimensionality reduction algorithm to qualitatively assess the accuracy of predictions locally within different regions of galaxy SED space.

### 4.1 Global comparisons

Given our ultimate goal of adding emission lines to mock spectra and using the results to characterize incompleteness in spectroscopic training sets, we are concerned with how well we are able to reproduce the detectability of a given line. For example, if the observed EW is ≈0.1 Å and we predict it instead to be ≈1 Å, it would still be predicted to be undetectable without long exposure times that would also have identified many continuum features for the same object. This is reflected in our choice of loss function, as explained in Section 3.1.

It should also be noted that the observed EWs are samples of intrinsic values perturbed by noise; as a result, we would not expect even an algorithm that perfectly predicts line EW from continuum flux to perfectly match the observations. Comparing line predictions to observations, especially at low EWs, must therefore account for these uncertainties. Definitively attributing differences between predicted and observed EWs to inaccurate predictions as opposed to noise in the observations is not possible without a perfect understanding of the latter, which is extremely difficult to attain.

*4.1.1 Summary statistics*

Due to catastrophic instrument or pipeline failures that might be present in early DESI data, we chose to quantitatively evaluate performance using the following three metrics that are robust to outliers:

(i) the **Spearman correlation coefficient**, defined as

$$\rho_s = \frac{\text{cov}[\text{R}(\text{EW}_{\text{pred}}), \text{R}(\text{EW}_{\text{obs}})]}{\sigma_{\text{R}(\text{EW}_{\text{pred}})}\sigma_{\text{R}(\text{EW}_{\text{obs}})}}, \tag{6}$$

---

[2] https://xgboost.readthedocs.io/en/stable/

**Table 1.** Spearman correlation coefficient, NMAD$\left[\frac{\Delta EW}{\sigma}\right]$, and fractional bias computed for the three methods presented in detail in Section 3, for each predicted line. The 'Masked Region' column indicates the wavelength windows within which the emission lines in the observed DESI spectra were masked to obtain the continuum. Our fiducial method of using a JAX-implemented simple neural network is shown as JAX-NN (Section 3.1). The most successful method previously used in the literature is identified as PCA (cf. Section 3.2.1). The results from predicting EWs from physical parameters are listed as PROVABGS (Section 3.2.2). The best values obtained for each line are indicated in bold. JAX-NN outperforms other methods in most cases, but only slightly; its main advantages are scalability via GPUs and differentiability.

| Line | Masked region (Å) | Spearman ($\rho_s$) | | | NMAD$\left[\frac{\Delta EW}{\sigma}\right]$ | | | Fractional bias ($F_b$) (per cent) | | |
|---|---|---|---|---|---|---|---|---|---|---|
| | | JAX-NN | PCA | PROVABGS | JAX-NN | PCA | PROVABGS | JAX-NN | PCA | PROVABGS |
| [O II]λ3726, λ3729 | 3718–3738 | **0.76** | **0.76** | 0.74 | 1.11 | 1.15 | **1.02** | −3.6 | **2.4** | −6.1 |
| H$\gamma$ | 4335–4349 | **0.74** | **0.74** | 0.72 | 0.89 | 0.89 | **0.92** | −5.2 | **−3.7** | −6.0 |
| H$\beta$ | 4856–4870 | **0.88** | **0.88** | 0.87 | 1.42 | **1.41** | 1.55 | **−1.4** | −3.4 | −10.0 |
| [O III]λ5007 | 5001–5015 | **0.77** | 0.75 | 0.75 | **1.30** | 1.35 | 1.32 | **−1.7** | 8.3 | 3.1 |
| H$\alpha$ | 6535–6579 | **0.92** | 0.91 | 0.91 | **4.90** | 5.00 | 5.00 | 0.7 | **−1.9** | −11.9 |
| [N II]λ6584 | 6578–6592 | **0.88** | 0.87 | 0.85 | **2.78** | 2.87 | 3.00 | **0.8** | −3.5 | −5.9 |
| [S II]λ6716 | 6711–6725 | **0.89** | 0.88 | 0.87 | **1.67** | 1.68 | 1.83 | **−0.1** | 0.8 | −6.3 |
| [S II]λ6731 | 6725–6739 | **0.87** | 0.86 | 0.85 | **1.45** | 1.46 | 1.56 | **−0.2** | 0.9 | −5.3 |

is a robust variation of the Pearson correlation coefficient, with R(X) representing the rank of each value of quantity X (e.g. 1 for the lowest value, 2 for the next, etc.);

(i) the **normalized median absolute deviation (NMAD)** of the residuals ($\Delta$EW) relative to the observed uncertainties ($\sigma_{EW}$), defined as

$$\text{NMAD}\left[\frac{\Delta EW}{\sigma_{EW}}\right] = 1.48 \times \text{median}\left[\left|\frac{\Delta EW}{\sigma_{EW}} - \text{median}\frac{\Delta EW}{\sigma_{EW}}\right|\right], \quad (7)$$

gives a measure of the spread between predicted and observed values relative to the estimated uncertainties in the observations, with a normalization such that it will converge to 1 for a Gaussian-distributed quantity if errors are purely due to noise that has been correctly estimated;

(i) and the **fractional bias**, defined as

$$F_b = \text{median}\left(\frac{EW_{pred}}{EW_{obs}} - 1\right). \quad (8)$$

These statistics were all calculated based upon raw EW values (converted from the predicted arcsinh(EW) values) for each of the methods described in detail in Section 3. The results are shown in Table 1, with the best values for a given quantity indicated in bold.

In an ideal case, the first two metrics will each be equal to one and the fractional bias will be zero. Spearman ($\rho_s$) values close to one indicate that the predictions are fairly accurate for every object, at least in the ranking sense (i.e. predicted EWs are largest for the objects with the greatest observed EW). Deviations of NMAD$\left[\frac{\Delta EW}{\sigma}\right]$ from one potentially indicate overfitting or underfitting; however, interpretation is complicated by the fact that the true uncertainties are not known. Finally, fractional bias ($F_b$) values close to zero indicate that there is no significant bias in the predictions.

For all the lines, we find strong correlations between predicted and observed values, with offsets that are comparable in order of magnitude to the EW uncertainties in the observations and biases that are small in most cases. This is true for all three of the methods shown in Table 1. It is no surprise that predicting the stronger lines (H$\alpha$, [N II]λ6584) is easier, with [O II] being an exception because it is a blended line that can be on top of a weak, uncertain continuum. The S/N of [O II] is also lower than the other strong lines, which is exacerbated by the fact that we use boxcar-average uncertainties, which should be significantly larger than the optimal line-profile uncertainties for blended lines. The Spearman statistic is worse for the lowest S/N lines ([O II] and H$\gamma$), most likely because noise causes scatter in the rank-ordering of observed fluxes. Generally, NMAD$\left[\frac{\Delta EW}{\sigma}\right]$ is larger than one, suggesting that the intrinsic scatter in the continuum–emission line relation is greater than the observed EW uncertainties (if those have been estimated accurately). This is most apparent for the stronger lines that have higher S/N. Biases from the continuum-based methods are small for most of the lines save H$\gamma$, which lies atop an absorption line that is sometimes significant. The negative fractional bias on H$\gamma$ could indicate that the absorption lines in the continuum under it are being overestimated, resulting in overestimated EWs from FastSpecFit (and comparatively underestimated predictions from our network). Biases from PROVABGS are in general worse. The weaker [N II]λ6548Å and [O III]λ4959Å lines can be easily estimated from the stronger ones that we predict by using the typical theoretical intensity ratios of [N II]λ6584/[N II]λ6548 ∼ 3 (Dojčinović, Kovačević-Dojčinović & Popović 2023) and [O III]λ5007/[O III]λ4959 ∼ 3 (Laker et al. 2022).

Overall, the three methods perform similarly, with the JAX-NN being scalable via GPUs, differentiable, easier to implement, and insensitive to the details of the spectrum. These advantages are important for adding realistic emission lines on to synthetic stellar continua. In addition, the PROVABGS method depends on the physical model used to extract these parameters, which adds an otherwise absent layer of complication; our JAX-NN method is less sensitive to systematically uncertain assumptions of galaxy SED evolution. The remainder of this section will only present analyses based upon our primary (JAX-NN) method.

### 4.1.2 Comparison of predicted versus observed equivalent widths

Plots comparing the EWs for each line predicted by the JAX-NN method to their observed EWs are shown in Fig. 3. The green contours[3] represent the distributions of predictions with added noise; this is done on an object-by-object basis by sampling values from a

---
[3]Contours appearing throughout this paper are obtained by using a Gaussian kernel density estimator from SciPy (Virtanen et al. 2020), with a bandwidth that is determined using Scott's Rule (Scott 2015). The quantiles of the resulting probability density functions are used to find different contour levels.

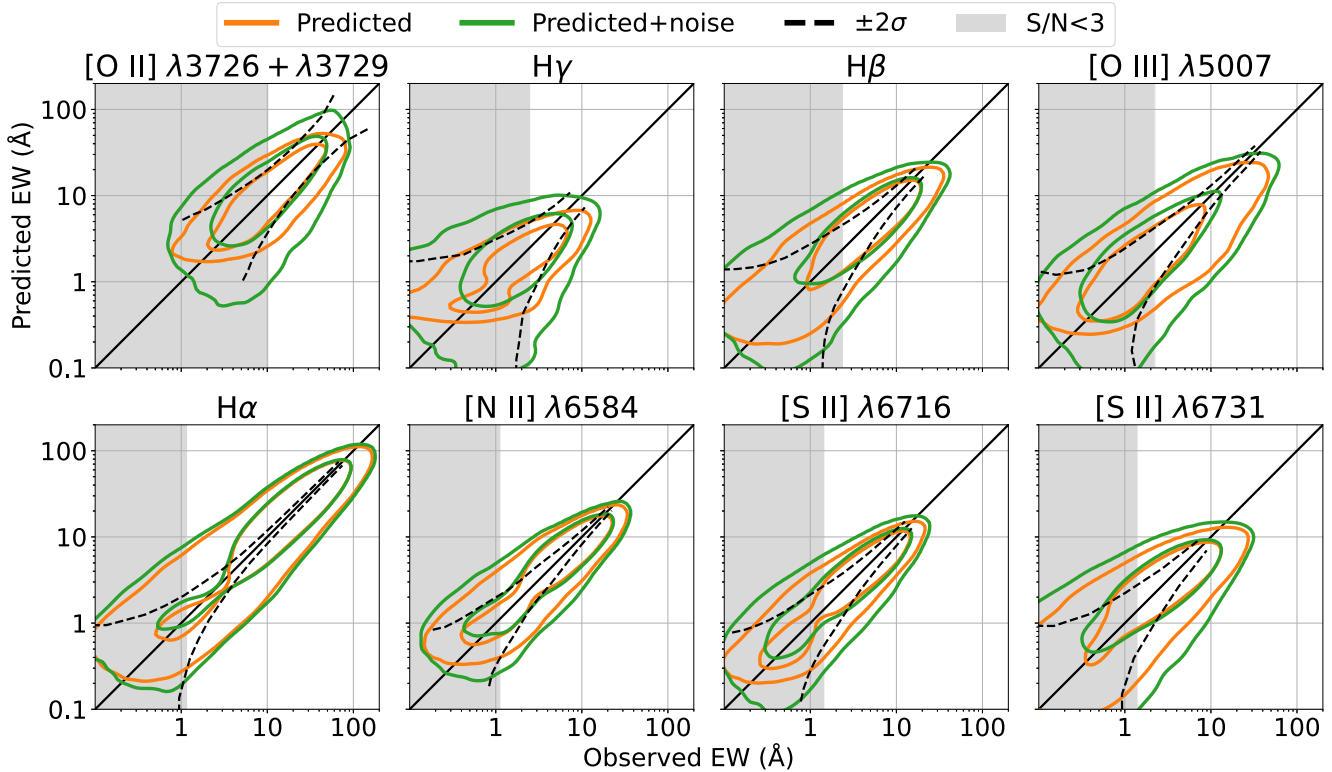

**Figure 3.** Comparisons of predicted versus observed EWs for all eight lines considered in this paper. The contours show regions within which 68 and 95 per cent of points lie, with (green) and without (orange) observational noise applied. The solid black line indicates the one-to-one line in each plot. The dashed black lines correspond to $\pm 2\sigma_{EW}$ regions assuming a log-normal distribution; they are obtained by binning the EWs (x-axis) and calculating within each bin the 2 per cent-trimmed mean of the EW uncertainties ($\sigma_{EW}$). The grey shaded areas indicate regions where on average the observed S/N < 3. There is a strong correlation between predicted and observed values, with scatter around the 1-to-1 line that is more than what would be expected from observational uncertainties alone, possibly due to the intrinsic variation in the continuum–emission line relation or due to misestimaion of errors.

Gaussian with mean equal to the prediction and standard deviation equal to the observed EW uncertainty for the corresponding object and line. For all the emission lines, the contours track the 1-to-1 line well, with ∼ 68 per cent of the points being tightly distributed around it. In the vast majority of cases, the predicted EWs differ from the observed EWs by an amount comparable in magnitude to the estimated measurement errors. For very low EWs (∼0.1) differences can reach an order of magnitude, but this is of little concern since these values have very low S/N and would correspond to a prediction that the given line is negligible for the purpose of redshift determination in any event.

The $\rho_s$ statistic in Table 1 is most sensitive to how well values track the one-to-one line in these plots. The NMAD$\left[\frac{\Delta EW}{\sigma}\right]$ statistic describes how spread the points are around this line relative to the dashed line which quantifies typical measurement errors. Finally, large values of the $F_b$ statistic would manifest as an asymmetric distribution of the contours about the one-to-one line. These plots also show that for observed EW values where the average observed S/N is less than three (grey regions), the spread in the predictions is on average less than what would be expected from the estimated measurement errors (dashed lines); the effects of adding noise to the predictions has correspondingly greater effect in this regime. This is particularly true for H$\gamma$, which is typically the weakest line amongst those we consider in this paper. The uncertainties at very large [O II] EWs increase mainly due to objects with weak, uncertain continua; such lines tend to be underpredicted. All the lines exhibit overpredictions at low values and underpredictions at high values to some degree, which is to be expected when applying regression methods to bounded data with non-negligible errors.

### 4.1.3 Cumulative distributions

Given our goal of predicting how often a given emission line would be detectable in a spectrum, it is important to check that the cumulative distributions of EWs for each line (i.e. how frequently a given line has EW below some value) are all realistic. Fig. 4 presents the results of this test. For all of the lines, we are able to accurately predict the fraction of objects that lie below a certain EW value, with only small deviations from the observed distributions after observational uncertainties are applied to the predictions. For example, the grey regions indicate values where the average S/N in the observed EW is less than three, and the predicted number of objects within this region is very similar to the observed one.

### 4.2 BPT diagrams: testing the reproduction of line ratios

In addition to testing the performance of our methods for each line individually, we also wish to investigate how well line ratios are preserved, given that we predict each line separately. We do this by constructing Baldwin–Phillips–Terlevich diagrams (BPT diagrams, Baldwin, Phillips & Terlevich 1981); we specifically focus on a line-ratio diagnostic diagram that uses the logarithm of the ratio of [N II] EW to H$\alpha$ EW ($\log[\frac{EW([N II])}{EW(H\alpha)}]$) on the x-axis and the logarithm of the ratio of [O III] EW to H $\beta$ EW (log[$\frac{\text{EW([O III])}}{\text{EW(H}\beta)}$]) on the y-axis. The location of a point on this diagram depends on the hardness of the ionizing radiation, making it useful for separating H II regions from AGNs. The dashed grey line in Fig. 5 (which we will refer to as 'Ke01') corresponds to the maximal ratios that can be produced by purely star-forming regions, as derived in Kewley et al. (2001). However, using data from the Sloan Digital Sky Survey (SDSS, York et al. 2000), Kauffmann et al. (2003) showed that most star-forming galaxies fall well below and to the left of this line, so they suggested a new empirical line, shown as the solid grey line (which we will refer to as 'Ka03') in the same figure. Objects to the left of the solid line are classified as star-forming; objects to the right of the dashed line are classified as AGN/LINERs; and objects in between are classified as composite (potentially including contributions from both star formation and other sources of ionization).

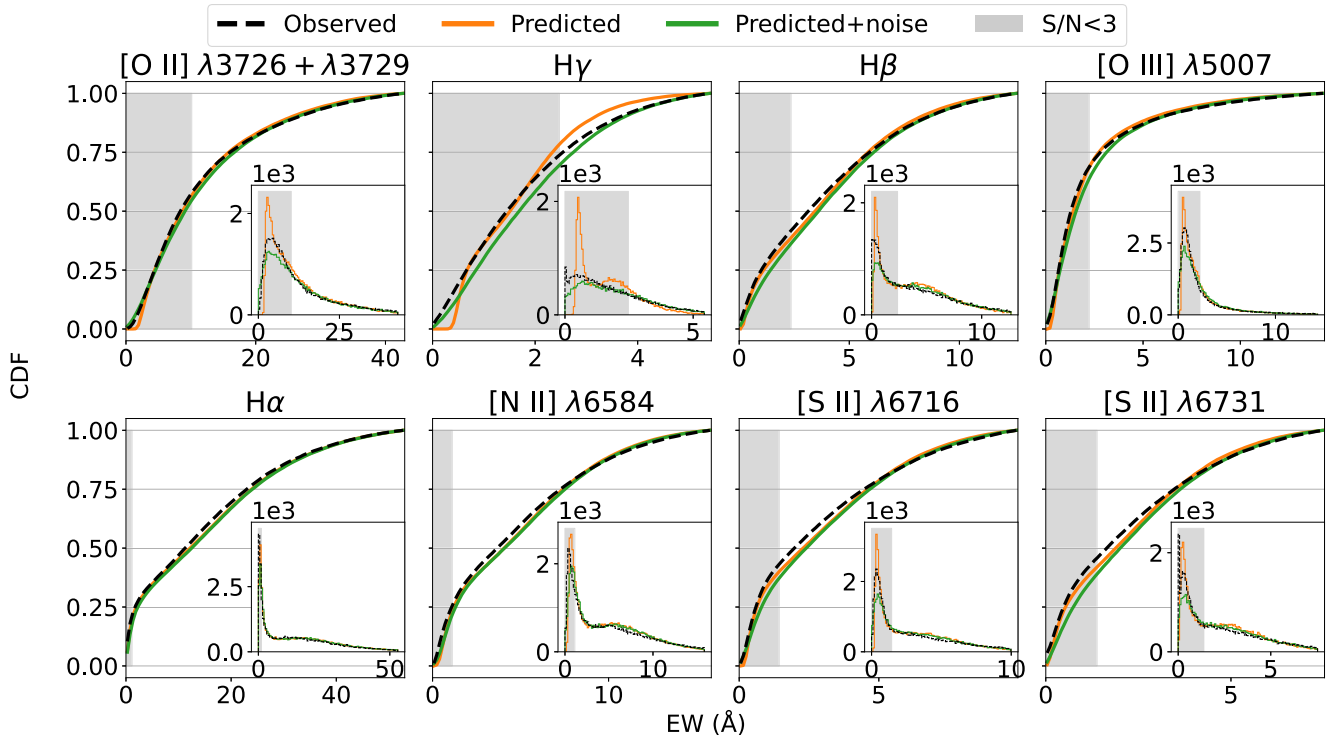

**Figure 4.** Cumulative distributions of the observed EWs (black), predicted EWs (orange), and predicted EWs with measurement errors applied (green). The grey shaded area shows the EW range within which the average observed S/N < 3 for a given line. The insets show histograms of the distribution of EW for each case. The predicted cumulative distributions with noise incorporated match the observed ones well, with discrepancies mainly in the grey region.

Fig. 5 shows the observed and reconstructed BPT diagrams for a variety of scenarios. The first diagram in the top panel shows all objects in the blind test set that have EW>0 for all the participating lines. The predicted distribution of points is tighter than the observed one (which is widened by observational errors) but displays the expected characteristics of a star-forming sequence and an AGN branch that connect at low [O III]/H $\beta$ values. Subsequent diagrams for which S/N cuts are applied illustrate how the observed distribution of objects in the BPT diagram depends greatly on the level of measurement errors. This is especially true when S/N > 3 is required for H $\beta$ and/or [O III], as those lines are weaker. Juneau et al. (2014) have shown similar effects using luminosity cuts on the emission lines.

The distributions of predicted points with noise added, shown in the bottom row of the figure, closely resemble the observed distributions. The added noise has much less impact in the [N II]/H $\alpha$ direction as both lines have higher S/N. However, both [O III] and H $\beta$ have non-negligible probabilities of being near zero when adding noise, which leads to heavy tails in the distribution of their ratio. Without noise, these tails are not recovered, mainly due to systematic overprediction of low EWs which can be seen in Fig. 3 (this is caused by doing regression with EWs that are bounded to be positive). In the limit where EW errors follow a Gaussian distribution, the ratio of two EWs will tend towards a Cauchy distribution when their mean values tend to zero (this will be exactly true if the means of the EW distributions are zero; cf. Ivezić et al. 2020).

Despite our JAX-NN method predicting EWs of each line separately, the distributions of line ratios and the relationship between different ratios appears to be preserved (within the context of the BPT diagram). This is not surprising, since the source of the ionizing radiation – the continuum that we use in our predictions—is what in large part what should determine these ratios.

### 4.2.1 Galaxies with weak emission

Previous research has shown that there is a significant population of weak line galaxies (WLGs; Fernandes et al. 2010) that are typically removed by S/N cuts when plotting BPT diagrams. The low-S/N H $\beta$ galaxies (WLG-H) mostly occupy the AGN/LINER region of the BPT; they are effectively removed when we apply a H $\beta$ S/N > 3 cut, as in the second column of Fig. 5. In contrast, low-S/N [O III] galaxies (WLG-O) mostly occupy the high metallicity end of the star-forming sequence (at the bottom of the V-shaped locus of galaxies in the figure); they are mostly removed by the [O III] S/N > 3 cut applied in the third column of the same figure. Those WLGs which are found in the AGN/LINER region represent either weak AGN or ionization driven by light from post-AGB stars in old stellar populations (Singh et al. 2013; Belfiore et al. 2016; Byler et al. 2019). For both the WLG-H and WLG-O populations we would expect to systematically overpredict EWs because of the lower bound of zero applied in the training sets. However, after realistic measurement errors are applied the distributions for such objects still match the observed distributions closely.

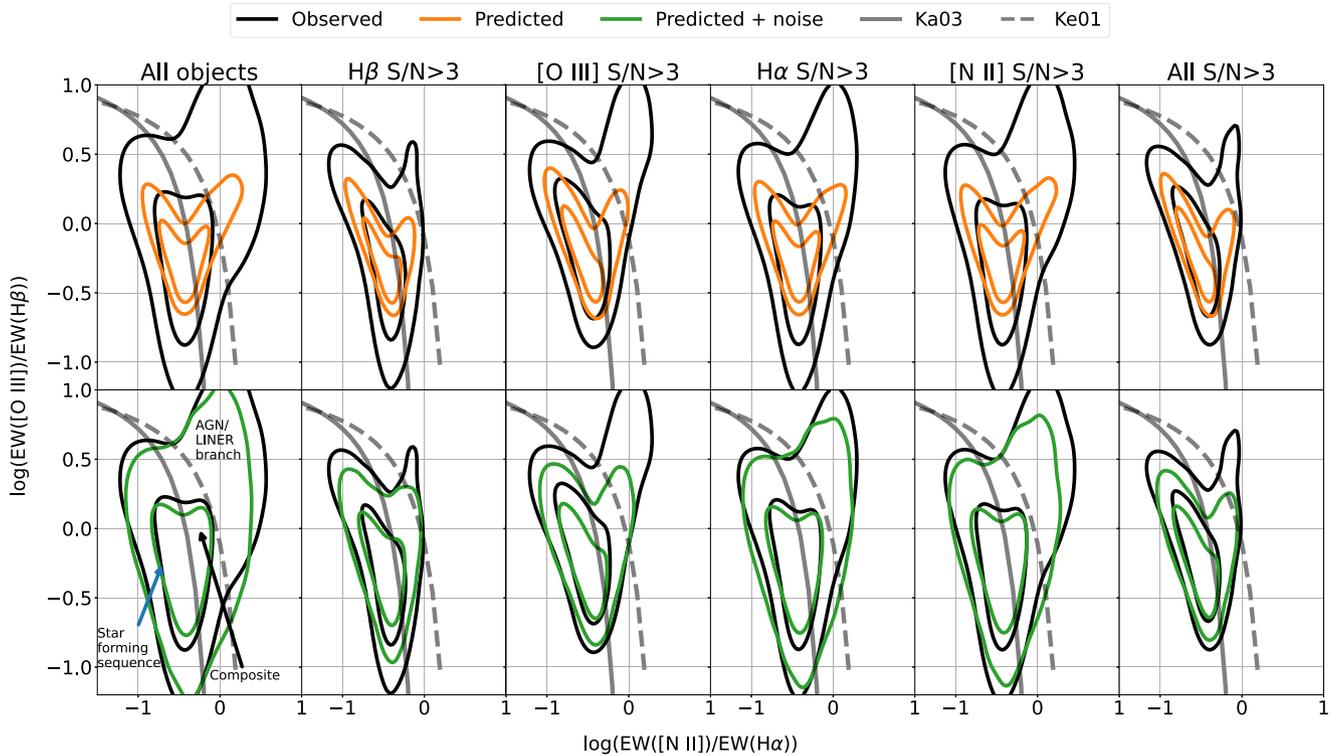

**Figure 5.** Comparison of observed BPT diagrams to predictions from our JAX-NN method. Note that we use EW([N II]$\lambda$6584) for the *x*-axis and EW([O III]$\lambda$5007) for the *y*-axis. The solid and dashed grey lines correspond to the Ka03 (Kauffmann et al. 2003) and Ke01 (Kewley et al. 2001) lines that can be used to separate the star-forming, composite, and AGN regions. The black contours represent boundaries within which 68 and 95 per cent of the observed points lie; orange and green contours show the corresponding regions for the values predicted from our network (top row) or for the predictions after realistic noise is added object-by-object (bottom row), respectively. The first column includes all the objects in the blind test set for which all relevant lines have EW>0 in both the observations and the noise-added predictions. Subsequent columns show only those objects which would remain in the set after an S/N > 3 cut on one individual line, while the last column requires S/N > 3 for all these lines (these cuts were applied to the observed and predicted values separately). The predicted (orange) distribution is tighter than the observed one (black); however, this discrepancy is much smaller when restricting to higher-S/N data, suggesting that much of the observed distribution of objects in this BPT diagram is due to noise. Indeed, the distribution of points when observational errors are added to our JAX-NN predictions is very similar to the observed distribution in every case.

In Fig. 6, we show the Veilleux & Osterbrock diagram (Veilleux & Osterbrock 1987), which is similar to the BPT diagram, but [N II] is replaced with [S II]$\lambda$6716 + $\lambda$6731. This diagram is another commonly used line-ratio diagnostic diagram that better separates LINERs from Seyferts. The solid grey line (Ka03) that separates star-forming galaxies from AGN is also from Kauffmann et al. (2003), and the dashed grey line (Ke06) was empirically determined by Kewley et al. (2006) to separate Seyferts from LINERs. We see similar trends as in Fig. 5, with noise playing an important role for reproducing extreme [O III]/H$\beta$ values. We emphasize that most of the galaxies in the Seyfert region of this diagram are not actually Seyferts, but galaxies with low S/N emission line strengths. Typically, emission line diagnostic diagrams are plotted after applying a S/N cut on the sample, which is shown in the last column of the figure.

### 4.3 2D-UMAP embedding of galaxy SEDs

Our analysis so far has focused on the global performance of our methods, assessed by the combined set of galaxies of all SEDs. However, ideally our method should not only reconstruct the properties of the population as a whole, but also be able to recover the distributions of EWs for galaxies at any point in the underlying space of galaxy SEDs. To assess this, we have employed Uniform Manifold Approximation and Projection (UMAP; McInnes et al. 2018b), a non-linear dimensionality reduction algorithm that preserves local topological structure, to produce a 2D representation of that underlying space and investigated the behaviour of our predictions across that simplified representation.

Specifically, we applied the UMAP algorithm to reduce the same set of features used to predict line EWs (i.e. a set of flux ratios and a single measure of luminosity) into two coordinates for each object. The UMAP mapping was trained with the complete H$\alpha$ EW>0 SV3 set (i.e. includes both test and blind-test sets) using the Python package UMAP (McInnes et al. 2018b). We used a nearest neighbour number of 30, minimum distance of zero, and default values for all remaining UMAP hyperparameters.

Fig. 7 shows the resulting 2D UMAP projections of the SV3 blind-test sets for the lines [O II], [O III], H$\alpha$, and [N II]. We emphasize that a single trained UMAP projection was used for all the panels; the training was done on the Hα SV3 set, and the projection was applied on the SV3 sets of the different lines shown. The 2D coordinates of the projected galaxy SEDs are clearly capable of separating star-forming galaxies with strong emission lines from passive galaxies with weak lines. This is consistent with low-dimensional embeddings of galaxy SEDs obtained with other methods, e.g. autoencoders (Portillo et al. 2020; Pat et al. 2022; Liang et al. 2023). The observed EWs follow a continuous trend between these two populations with some scatter; there are galaxies that have strong emission but whose neighbours in the UMAP have weak emission, and other galaxies that have weak emission but are surrounded by objects with strong emission. This is presumably due to a combination of observational uncertainties and intrinsic scatter in the continuum–emission line relationship.

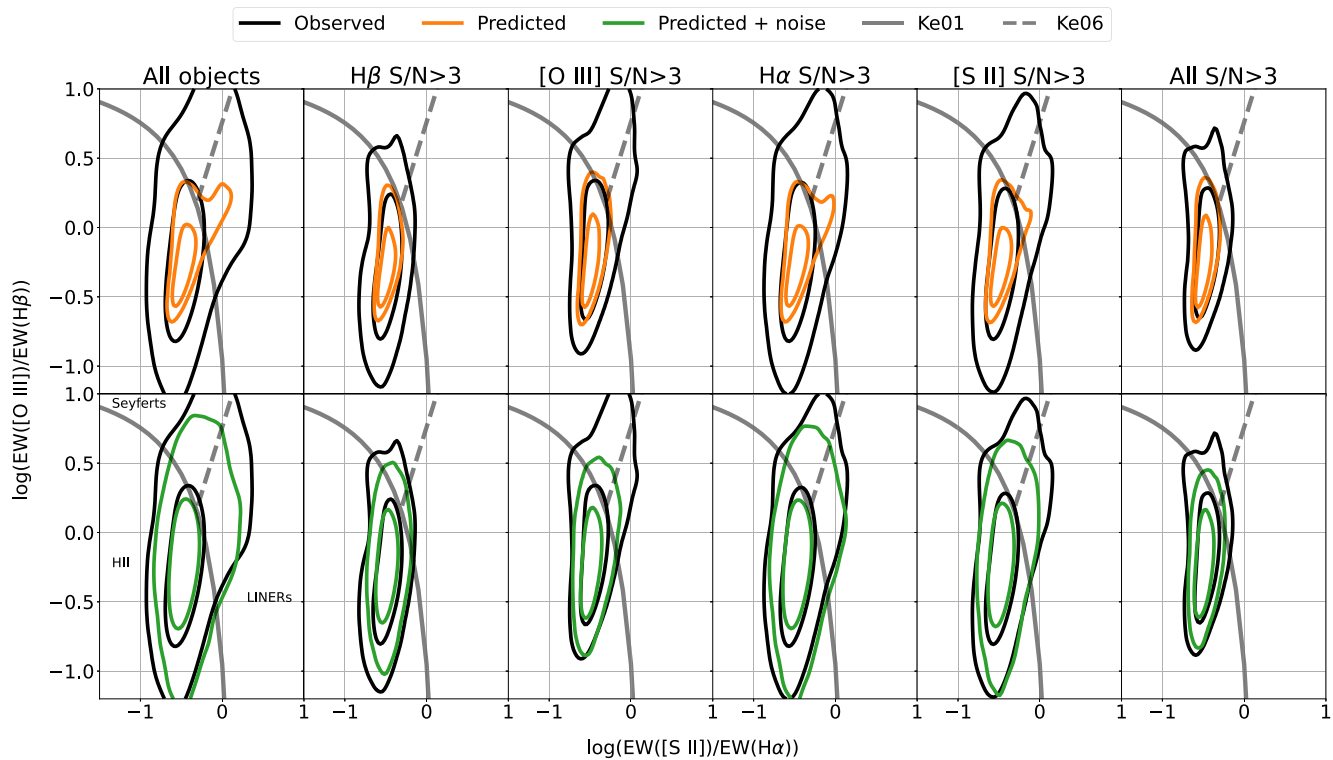

**Figure 6.** Comparison of observed Veilleux and Osterbrock diagrams (Veilleux & Osterbrock 1987) to predictions from our JAX-NN method. We use EW([S II]λ6716) + EW([S II]λ6731) for the *x*-axis and EW([O III]λ5007) for the *y*-axis. The solid grey (Ke01, Kauffmann et al. 2003) and dashed grey (Ke06, Kewley et al. 2006) lines separate star-forming galaxies, Seyferts, and LINERs. The contours and the different columns of S/N cuts are the same as in Fig. 5, except that [N II] is replaced with [S II].

Two distinct regions show departures from the dominant trend. Through visual inspection, we found that objects at the middle right of the UMAP locus generally exhibit strong, broad AGN emission, even though they should be excluded by our requirement that all spectra be best fit by galaxy templates (cf. Section 2.3). In contrast, objects at the bottom right of the locus generally correspond to galaxies with very small flux in the bluest bin, likely due to a combination of the lower throughput, higher sky background in bright time, and greater calibration issues at the bluest end of the DESI spectrographs (Guy et al. 2023).

The predicted EWs from JAX-NN trace the continuous overall trend seen for the observed EWs, but with much smaller scatter at a given point in UMAP space. This is to be expected since for a given neighbourhood in SED space, minimizing the training loss should drive our neural network to reproduce the average EW (since an average minimizes the sum of the squares of deviations). Interestingly, the predictions with added noise match the local scatter of the observed EWs better but not perfectly. This suggests that there remains some intrinsic scatter in the relationship between continuum spectrum and emission line EWs that is not captured by our neural network.

### 4.3.1 BPT classes in UMAP space

To provide a better understanding of how galaxy populations correspond to position in UMAP space, we have classified all objects as either purely star-forming (SF), composite (SF and/or AGN/LINER), or AGN/LINER, based on their locations in the BPT diagram (see Section 4.2). Objects to the left of the solid grey line in Fig. 5 (Ka03; Kauffmann et al. 2003) were classified as purely SF, objects to the right of the dashed grey line (Ke01; Kewley et al. 2001) were classified as AGN/LINER, and objects in between were classified as composite.

Fig. 8 shows the 2D UMAP projection colour coded by the resulting BPT-based classifications. Much like the observed EWs used in Fig. 7, the BPT classes follow a continuous trend with some scatter. Most of the weak-line galaxies are either weak AGN or LINERs which can exhibit emission from gas that is radiated by old stellar populations (post-AGB stars) (Singh et al. 2013; Belfiore et al. 2016; Byler et al. 2019). Echoing the trend from weak to strong emission lines across UMAP space, galaxies also transition from predominantly AGN/LINER to composites to star forming. There is significant mixing between the observed AGN/LINER and composite population, due to some combination of observational errors mixing classes and intrinsic scatter in the relationship between continuum and line emission. This is exacerbated by the fact that the BPT classifications do not definitively correspond to underlying physical distinctions. The composite population could include objects whose emission is sourced by varying combinations of AGN, young stars, and post-AGB stars; and even objects to the right of the Ke01 line which are classified as AGN/LINERs can have some emission coming from H II regions (Agostino et al. 2021).

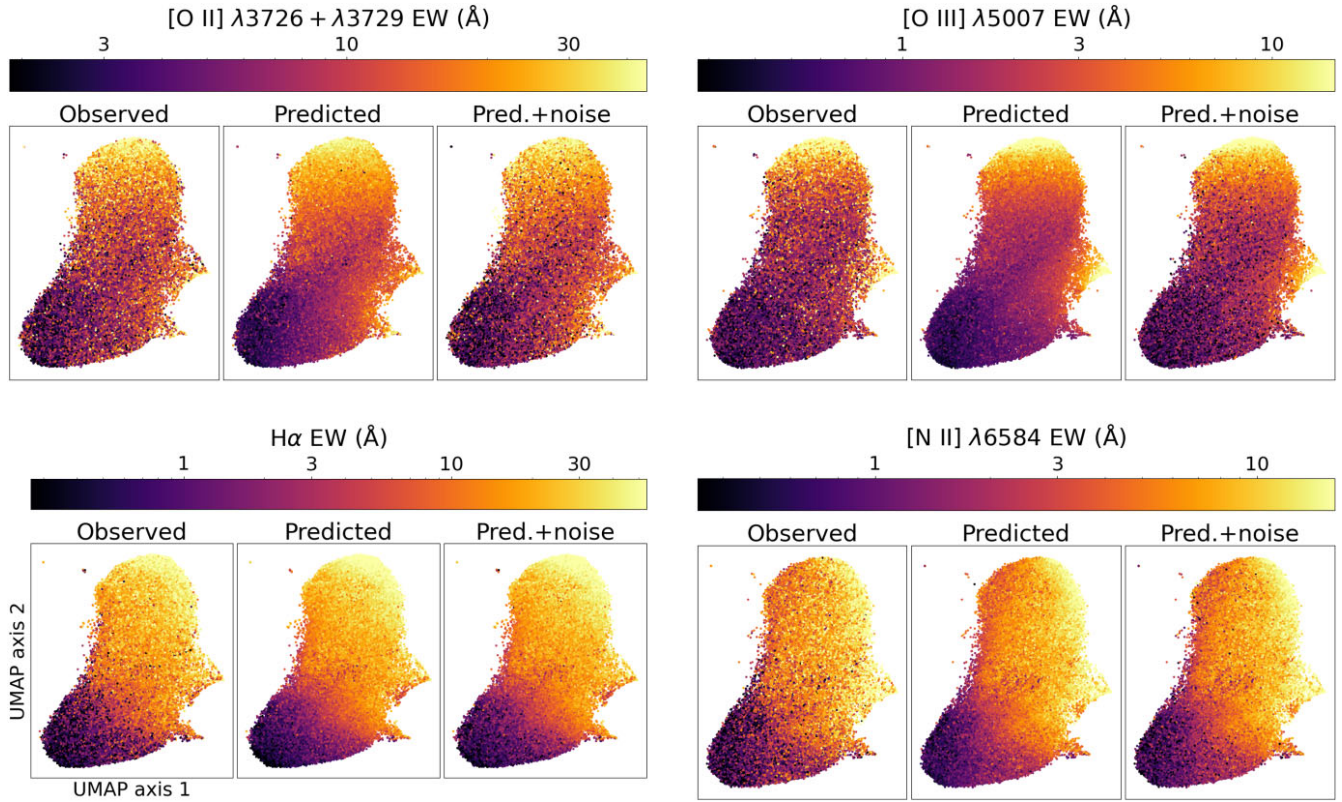

**Figure 7.** A test of line prediction accuracy across the observed parameter space (continuum flux ratios and luminosity around 6250 Å). The panels show positions of SV3 blind test set objects in a 2D UMAP embedding which remaps the same quantities used to predict emission lines. We emphasize that the same projection is used for all the panels, and the resulting 2D space does not have physical meaning. Each point is colour coded according to the EW for a given emission line. We plot the observed EWs, the EWs predicted by JAX-NN, and the predictions with noise added in separate panels. The corresponding plots for the remaining lines listed in Table 1 but not shown here closely resemble the results for H$\alpha$ (up to a normalization of EWs). Predicted EWs only follow the continuous trend of observed EWs. Some of the observed variation around this trend is recovered when noise is added to the predictions, but the remaining variation, which is due to intrinsic scatter in the continuum–emission line relation, is not captured by our method.

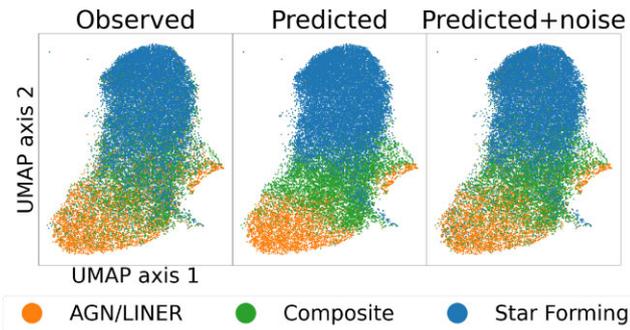

**Figure 8.** Same as Fig. 7, but colour coded with BPT class rather than line EW. The plotted sample corresponds to the H$\alpha$ blind test set used for the lower left panel of that figure. As with the emission line EWs, the predicted BPT classes follow the average continuous trend of the observed ones, and adding noise to the emission line predictions before determining the BPT class recovers much of the observed scatter.

The BPT diagram can also be thought of as a low-dimensional embedding of galaxy SEDs but derived from emission lines instead of continuum shape. Indeed, the UMAP embedding of Fig. 7 traces the V-shape of the BPT (Fig. 5). At the highest values of the UMAP $y$-axis, galaxies have strong emission lines (including strong [O III]), indicating that these are low-metallicity star-forming galaxies that occupy the top-left region of the star-forming sequence in the BPT diagram. At lower $y$ values, [O III] and [O II] quickly become negligible, whereas H$\alpha$ and [N II] are still relatively strong. These are high-metallicity star-forming galaxies, and they occupy the bottom of the V-shape in the BPT diagram. Near the Ka03 line in the BPT, star-forming galaxies are mixed with and transition to composite populations, which eventually transition to AGN/LINERs or Seyferts past the Ke01 line. This is also reflected in the UMAP; intermediate $y$-values that correspond to high-metallicity star-forming galaxies transition to composites and weak AGN/LINERs at lower $y$-values, or they transition to composites and Seyferts at intermediate $y$-values and high $x$-values, as indicated by stronger [O III] and [N II] emission.

### 4.4 Summary of results

To summarize, we have found that the predicted EWs from our neural network are strongly correlated with the observed EWs, both when the galaxy population is considered as a whole and when only objects of similar SED are compared. Differences are primarily attributable to uncertainties in the observed emission line equivalent widths, but we also find some evidence for intrinsic scatter in the continuum–emission line EW relation at a subdominant level; i.e. for a given continuum shape a range of EWs would be observed even if random measurement errors were zero. Given that our network is only provided the former, it is unable to learn the observed variation in EWs for a fixed SED. Adding observational uncertainties to our predictions is critical for reproducing the observed distributions of EWs and EW ratios in the BGS sample; this is particularly important for reproducing the tails of the line ratio distributions (especially for [O III]/H$\beta$) which are dominated by measurements with low *S/N*.

Our neural network is able to predict the EWs of strong AGN present in the BGS sample, which occupy a separate region in UMAP space (as seen in Figs 7 and 8). This suggests that it may be possible to use our method to also predict emission lines for quasars. As a simple test, we checked our predictions for roughly ∼80 galaxies in our test set that have [Ne V]$\lambda3346$ S/N > 3, which indicates an AGN contribution (Reefe et al. 2023). We found that the H$\beta$, H$\alpha$, [N II], and low and high [O III]$\lambda5007$ predictions are reliable. however, intermediate [O III] values (10–80Å) are underpredicted. Presumably, for such cases, the AGN's contribution to the continuum is negligible, and so our neural network, given only the stellar continuum, predicts an [O III] EW that does not include the AGN's contribution. This is not too concerning, since the line is still predicted to be detectable for the purposes of redshift measurement. For applications that involve AGN in general, further testing is required and is beyond the scope of this paper (and not relevant for our primary goals of predicting whether objects will have strong enough line emission to detect in a spectrum).

## 5 SUMMARY, CONCLUSIONS, AND FUTURE WORK

### 5.1 Overview

We have trained a JAX-implemented Neural Network on DESI BGS spectra to predict the equivalent widths of eight strong optical emission lines from a galaxy's continuum spectrum. A measure of luminosity at roughly 6250 Å together with 11 flux ratios between successive windows within the rest-frame optical continua were used as inputs for the predictions. In Section 4.1, we demonstrated through comparisons of predicted and observed EW distributions that our network is able to produce a realistic distribution of line strengths. Some discrepancies are present, attributable to some combination of noise in the observations and intrinsic scatter in the relation between emission line and continuum properties.

In Table 1, we have compared the results of the JAX-NN method to predicting EWs from PCA coefficients using local weighted linear regression, as was done in Beck et al. (2016), as well as to predicting EWs from a set of physical parameters obtained from the PROVABGS catalogue. The three methods performed similarly, with our method typically doing slightly better. However, unlike alternatives, the JAX-NN is scalable via GPUs, simple to implement, insensitive to the details of the continuum spectrum, and differentiable; we therefore have focused exclusively on JAX-NN in the remainder of the paper.

In Section 4.2, we constructed BPT diagrams from our predicted EWs and compared to equivalent plots from observations in order to test whether the predictions preserve relationships between lines (e.g. whether line ratios match physical ones). We found that the predicted distribution of points exhibited a star-forming sequence and an AGN branch connected in a V-like shape, as is also true for observed samples. However, adding noise to the predictions based on the estimated observational errors for each object was essential for recovering the full range of observed values, especially for the [O III]/H$\beta$ ratio which relies on weaker lines. It is of course not surprising that a machine learning algorithm would predict a tighter distribution than what is observed when errors are not accounted for; the impact is particularly large for ratios of quantities measured with low signal-to-noise ratio as EW values which are low due to scatter will lead to heavy tails in the distribution of ratios.

Both Sections 4.1 and 4.2 assessed the performance of our method in a global sense; i.e. considering objects of all properties together. In order to test whether our methods are in fact effective everywhere within the range of different galaxy SEDs, we applied the UMAP dimensionality reduction method on the feature space that was used to predict EWs (i.e. the set of flux ratios and a luminosity). In Section 4.3, we use this to demonstrate that we are able to predict EWs well across the full space, though there is evidence for a subdominant intrinsic scatter in the distribution of line EWs at fixed continuum SED.

### 5.2 Applications of the JAX-NN algorithm

The scalability and differentiability inherent to the JAX-NN method make it straightforward to integrate this model with the DSPS population synthesis code (Hearin et al. 2023). Since it is implemented in JAX, DSPS offers a differentiable, GPU-optimized alternative to traditional SPS codes. Because our model is also implemented in JAX, it enables adding empirically motivated emission line strengths to stellar continua synthesized by DSPS while retaining its advantages of differentiability and scalability. A pipeline which combines the two codes can speed up both forward-modelling galaxy populations and inferring galaxy properties from observations, while including realistic emission line distributions.

Currently, DSPS incorporates photoionization-based emission lines by adding self-consistent nebular emission to SSP templates, following Byler et al. 2017. However, using such templates in a computationally efficient way involves making strong and likely unrealistic assumptions. That approach could be compared to our empirical method by generating two DESI BGS-like synthetic catalogues: one with photoionization-based emission lines and one with emission lines predicted by our method. The resulting distributions of line strengths from both catalogues can be compared directly to the observed distributions in DESI BGS, much as we have already done for JAX-NN here.

If the photoionization-predicted lines prove sufficiently realistic in this scenario, then we should be able to have some trust in them for different galaxy populations (e.g. at higher redshifts). Our methods should not extrapolate beyond the limits of the training set; however, there is significant evidence that the range of galaxies observed at low $z$, including populations of highly star-forming dwarfs which should be represented within BGS, include objects with similar SEDs to even the most extreme objects observed at high redshift (e.g. Mingozzi et al. 2023). This suggests that our methods (possibly supplemented with extra parameters that can be used to forward-model evolution in the line-continuum relationship) should still be sufficient to produce mock catalogues with at least somewhat realistic spectroscopic incompleteness. An alternative approach would be to use our results to tune photoionization parameters for DSPS if they prove lacking, as extrapolating to different populations of galaxies should be better-behaved with a physical model as opposed to a neural network. The extrapolation to higher redshifts can be tested on ≈200 extremely metal-poor low-redshift galaxies that have been identified in DESI early data (Zou et al. 2024); thousands of such objects will be eventually observed by DESI, especially through the LOW-Z secondary target program (Darragh-Ford et al. 2023).

Our motivation for this work has been to ultimately improve the modelling of the effects of incompleteness in the spectroscopic training and calibration sets used for photometric redshifts. An important source of incompleteness stems from the fact that objects in real training sets should have at least two strong spectral features, typically emission lines, to enable a secure redshift measurement. Hartley et al. (2020) have shown using a simple mock catalogue that the bias in photo-$z$ predictions resulting from this incompleteness is significantly larger than the requirements for Rubin Observatory LSST dark energy inference uncertainties not to be dominated by photo-$z$ systematics (Mandelbaum et al. 2018). A more sophisticated mock catalogue could be used to characterize (and potentially develop methods to mitigate) the bias caused by this incompleteness.

Given a realistic distribution of galaxy physical parameters, our method, in conjunction with DSPS, provides a way of synthesizing spectra with realistic emission lines. With such a catalogue, it will be possible to model which observed galaxies should yield accurate redshift measurements given a particular spectroscopic instrument (such as DESI) and exposure time. This would effectively characterizse the probability of a successful redshift measurement as a function of a galaxy's colours, magnitude, and redshift. If such probabilities are well-understood, it will be possible to potentially improve photo-$z$ estimates by appropriate re-weighting (Newman & Gruen 2022). A DSPS pipeline with empirically motivated emission lines can also be used to potentially improve the inference of galaxy properties from measurements, combining information from both the emission lines and continuum features.

We note that our network was trained on DESI Early Release data Collaboration (2023b). When year 1 data is released, it would be possible to use at least an order of magnitude more objects for training a more complicated, deeper neural network that could better capture the continuum–emission line relation. However, these spectra will have lower signal-to-noise ratio than the SV1 observations provide, so it remains to be seen whether this will improve performance.

## ACKNOWLEDGEMENTS

We would like to thank Andy Connolly, Dalya Baron, Rachel Bezanson, and Viviana Acquaviva for fruitful discussions.

The efforts of J. A. Newman and A. Khederlarian were supported by grant DE-SC0007914 from the U.S. Department of Energy Office of Science, Office of High Energy Physics.

This material is based upon work supported by the U.S. Department of Energy (DOE), Office of Science, Office of High-Energy Physics, under Contract No. DE-AC02-05CH11231, and by the National Energy Research Scientific Computing Center, a DOE Office of Science User Facility under the same contract. Additional support for DESI was provided by the U.S. National Science Foundation (NSF), Division of Astronomical Sciences under Contract No. AST-0950945 to the NSF's National Optical-Infrared Astronomy Research Laboratory; the Science and Technology Facilities Council of the United Kingdom; the Gordon and Betty Moore Foundation; the Heising-Simons Foundation; the French Alternative Energies and Atomic Energy Commission (CEA); the National Council of Science and Technology of Mexico (CONACYT); the Ministry of Science and Innovation of Spain (MICINN), and by the DESI Member Institutions: https://www.desi.lbl.gov/collaborating-institutions. Work done at Argonne National Laboratory was supported under contract no. DE-AC02-06CH11357. Any opinions, findings, and conclusions or recommendations expressed in this material are those of the author(s) and do not necessarily reflect the views of the U. S. National Science Foundation, the U. S. Department of Energy, or any of the listed funding agencies.

The authors are honored to be permitted to conduct scientific research on Iolkam Du'ag (Kitt Peak), a mountain with particular significance to the Tohono O'odham Nation.

A. Khederlarian thanks the Legacy Survey of Space and Time Discovery Alliance (LSST-DA) Data Science Fellowship Program, which is funded by LSST-DA, the U.S National Science Foundation (NSF) Cybertraining Grant #1829740, the Brinson Foundation, and the Moore Foundation; his participation in the program has benefited this work.

A. Khederlarian thanks Argonne National Lab, the Ludwig Maximilian University of Munich, and the Vatican Observatory for providing the support and opportunity to present this work and discuss with colleagues.

*Softwares:* NUMPY (Harris et al. 2020), MATPLOTLIB (Hunter 2007), SCIPY (Virtanen et al. 2020), PANDAS (McKinney 2010), ASTROPY (Astropy Collaboration 2022), UMAP (McInnes et al. 2018b), JAX (Bradbury et al. 2018), SKLEARN (Pedregosa et al. 2011), HEALPY (Zonca et al. 2019).

## DATA AVAILABILITY

The data used to produce all results in this paper are available to the public as part of the DESI Early Release data.[4] The parameters of our JAX-NN network and documentation of how to use it are in a Github repository.[5] The data used to make the plots in this paper can be found on Zenodo.[6]

---

[4] https://data.desi.lbl.gov/doc/
[5] https://github.com/ashodkh/emission-lines
[6] https://zenodo.org/records/10815266

[1]*Department of Physics and Astronomy and PITT PACC, University of Pittsburgh, Pittsburgh, PA 15260, USA*
[2]*Department of Physics and Astronomy, Siena College, 515 Loudon Road, Loudonville, NY 12211, USA*
[3]*HEP Division, Argonne National Laboratory, 9700 South Cass Avenue, Lemont, IL 60439, USA*
[4]*NSF's NOIRLab, 950 N Cherry Avenue, Tucson, AZ 85719, USA*
[5]*Fakultät für Physik, Ludwig-Maximilians-Universität München, Universitäts-Sternwarte, Scheinerstr. 1, D-81679 München, Germany*
[6]*Department of Astrophysical Sciences, Princeton University, Princeton, NJ 08544, USA*
[7]*Institute of Cosmology and Gravitation, University of Portsmouth, Dennis Sciama Building, Portsmouth PO1 3FX, UK*
[8]*Lawrence Berkeley National Laboratory, 1 Cyclotron Road, Berkeley, CA 94720, USA*
[9]*Boston University, 590 Commonwealth Avenue, Boston, MA 02215, USA*
[10]*Department of Physics & Astronomy, University College London, Gower Street, London WC1E 6BT, UK*
[11]*Instituto de Física, Universidad Nacional Autónoma de México, Cd. de México C.P. 04510, México*

[12]*Kavli Institute for Particle Astrophysics and Cosmology, Stanford University, Menlo Park, CA 94305, USA*
[13]*SLAC National Accelerator Laboratory, Menlo Park, CA 94305, USA*
[14]*Berkeley Center for Cosmological Physics, University of California, Berkeley, 110 Sproul Hall #5800 Berkeley, CA 94720, USA*
[15]*Departamento de Física, Universidad de los Andes, Cra. 1 No. 18A-10, Edificio Ip, CP 111711, Bogotá, Colombia*
[16]*Observatorio Astronómico, Universidad de los Andes, Cra. 1 No. 18A-10, Edificio H, CP 111711 Bogotá, Colombia*
[17]*Institut d'Estudis Espacials de Catalunya (IEEC), E-08034 Barcelona, Spain*
[18]*Institute of Space Sciences, ICE-CSIC, Campus UAB, Carrer de Can Magrans s/n, E-08913 Bellaterra, Barcelona, Spain*
[19]*Department of Physics, Southern Methodist University, 3215 Daniel Avenue, Dallas, TX 75275, USA*
[20]*Departament de Física, Serra Húnter, Universitat Autònoma de Barcelona, E-08193 Bellaterr, Barcelona, Spain*
[21]*Institut de Física d'Altes Energies (IFAE), The Barcelona Institute of Science and Technology, Campus UAB, E-08193 Bellaterra, Barcelona, Spain*
[22]*Institució Catalana de Recerca i Estudis Avançats, Passeig de Lluís Companys, 23, E-08010 Barcelona, Spain*
[23]*Department of Physics and Astronomy, University of Sussex, Brighton BN1 9QH, UK*
[24]*Department of Physics & Astronomy, University of Wyoming, 1000 E. University, Dept. 3905, Laramie, WY 82071, USA*
[25]*National Astronomical Observatories, Chinese Academy of Sciences, A20 Datun Rd., Chaoyang District, Beijing 100012, P.R. China*
[26]*Space Sciences Laboratory, University of California, Berkeley, 7 Gauss Way, Berkeley, CA 94720, USA*
[27]*Instituto de Astrofísica de Andalucía (CSIC), Glorieta de la Astronomía, s/n, E-18008 Granada, Spain*
[28]*Department of Physics, Kansas State University, 116 Cardwell Hall, Manhattan, KS 66506, USA*
[29]*Department of Physics and Astronomy, Sejong University, Seoul, 143-747, Korea*
[30]*CIEMAT, Avenida Complutense 40, E-28040 Madrid, Spain*
[31]*Department of Physics, University of Michigan, Ann Arbor, MI 48109, USA*

This paper has been typeset from a T<sub>E</sub>X/L<sup>A</sup>T<sub>E</sub>X file prepared by the author.